\documentclass{egpubl}

 \JournalSubmission    
 \electronicVersion 

\ifpdf \usepackage[pdftex]{graphicx} \pdfcompresslevel=9
\else \usepackage[dvips]{graphicx} \fi

\PrintedOrElectronic

\usepackage{t1enc,dfadobe}

\usepackage{egweblnk}
\usepackage{cite}

\usepackage{algorithmic}
\usepackage{algorithm}
\usepackage{amssymb,amsmath}
\usepackage{wrapfig}

\usepackage{caption}
\title{Shapes In A Box - Disassembling 3D objects for efficient packing and fabrication}
\author[Marco Attene]
{Marco Attene
\\
Istituto di Matematica Applicata e Tecnologie Informatiche, Consiglio Nazionale delle Ricerche - Genova (Italy)
}

\begin{document}

\maketitle

\begin{abstract}
Modern 3D printing technologies and the upcoming mass-customization paradigm call for efficient methods to produce and distribute arbitrarily-shaped 3D objects.
This paper introduces an original algorithm to split a 3D model in parts that can be efficiently packed within a box, with the objective of reassembling them after delivery.
The first step consists in the creation of a hierarchy of possible parts that can be tightly packed within their minimum bounding boxes.
In a second step, the hierarchy is exploited to extract the (single) segmentation whose parts can be most tightly packed.
The fact that shape packing is an NP-complete problem justifies the use of heuristics and approximated solutions whose efficacy and efficiency must be assessed.
Extensive experimentation demonstrates that our algorithm produces satisfactory results for arbitrarily-shaped objects while being comparable to ad-hoc methods when specific shapes are considered.

\begin{classification} 
\CCScat{Computer Graphics}{I.3.5}{Computational Geometry and Object Modeling}{Hierarchy and geometric transformations}
\end{classification}

\end{abstract}



\section{Introduction}
3D fabrication is becoming so important that specific international conferences (\cite{ins3d2013}) and schools \cite{schoolfab2013} are being organized to discuss  this technology and its impact on both academia and industry. The increasing popularity of 3D printers goes exactly in the direction of the mass-customization paradigm evisaged by most modern market trends, and some experts already consider these devices as triggers for the next industrial revolution. Unfortunately, 3D printers represent a relatively new technology, and printing complex and articulated 3D objects \cite{cali2012} easily requires to face problems such as physical robustness of the printed prototype \cite{stava2012} and maximum size of a printable object \cite{luo2012}.
Furthermore, delivering a customized printed object (e.g. to a customer) has a cost that grows as the size of the ``pack'' grows.
Thus, it is important to investigate how a 3D model can be split into easily printable parts that can eventually be tightly packed in a box and reassembled at destination.

\begin{figure*}
   \includegraphics[width=\linewidth]{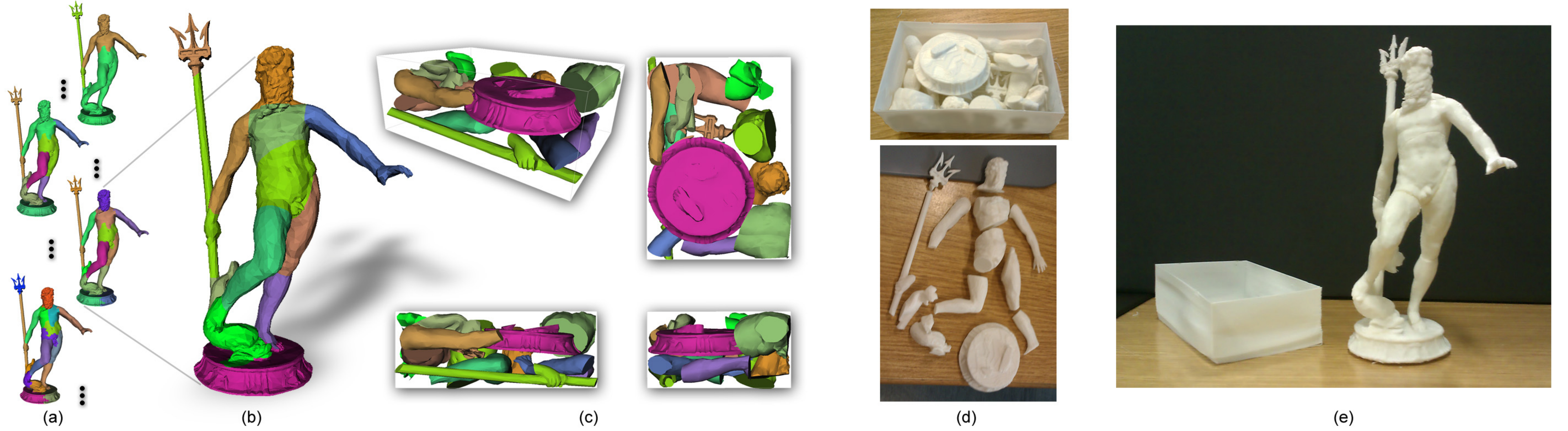}
   \caption{Starting from a tetrahedral mesh, the split-and-pack algorithm presented in this article produces a hierarchical segmentation (a), extracts an appropriate resolution out of the hierarchy (b), and computes a tight arrangement of the parts within a container (c) to be 3D-printed and shipped. At destination the printed parts can then be extracted from their pack (d) and eventually reassembled (e).}
   \label{fig:teaser}
\end{figure*}

This paper introduces an original algorithm to perform the aforementioned operation automatically. Specifically, the algorithm splits an input 3D model into a set of simple parts, calculates the size of a box that can contain all the parts and determines an arrangment of the parts in the box. While developing the algorithm, we took into account the following two desireable features:
\begin{itemize}
\item the eventual box must be as small as possible (delivering large boxes is expensive);
\item the number of parts must not be too large (the object must be reassembled in a reasonable time).
\end{itemize}
When splitting an object for 3D printing, other desireable conditions exist but are often in conflict with each other \cite{luo2012} even without considering the packing problem. Thus, while attempting to find a good compromise, herewith we focus on the packing issue for which a solution is necessary, while we point the reader to the aforementioned previous works for applications having different priorities.

\subsection{Algorithm overview}
Our approach is founded on the integration of two previously-studied problems, namely, shape segmentation and the bin packing problem. Specifically, in a first step we segment the object into parts using a criterion that allows the parts themselves to be tightly packed. Then, in a second step we compute a roto-translation (i.e. a proper rigid transformation) for each of the parts, compute their overall minimum bounding box, and measure the \emph{packing efficiency} as the ratio between the object volume and the box volume. Clearly, the packing efficiency is likely to increase as the number of parts increases but, as we mentioned, the number of parts cannot grow indefinitely. Thus, it is necessary to find a good compromise.

\subsection{Summary of contributions}
This article brings several contributions to areas such as computer graphics, shape analysis and operational research. Our original contributions can be summarized as follows:
\begin{itemize}
\item A novel split-and-pack integrated approach to find a tradeoff between the number of parts and the packing efficiency;
\item A new segmentation algorithm that produces hierarchies of box-like parts;
\item A new packing algorithm to properly place the parts into a box.
\end{itemize}

When FDM printers are used (see Sec. \ref{sec:reassembling}), packing the parts before printing may lead to a saving of both printing time and costs due to a smaller usage of support material. However, we point out that this is not the objective of this paper and, as discussed in Sec. \ref{sec:results}, we focus on the minimization of the final box volume.

\section{State of the art}
Our algorithm attempts to partition the object so that the resulting parts are suitable for packing, and then tries to arrange the parts so that they can be enclosed within a small box. Broadly speaking, algorithms to partition a 3D object can be classified as shape segmentation techniques: within this family, some algorithms compute hierarchical segmentations to recursively subdivide the object into smaller and smaller parts. Hierarchical segmentation is particularly useful for collision detection applications, where the space occupied by the object is typically subdivided using different sorts of bounding containers (e.g. axis-aligned boxes, oriented boxes, convex hulls, minimum-volume boxes, ...). Tight arrangements of sets of geometric objects are mostly studied in operational research, where various formulations of packing and related problems are studied.

Thus, the previous work we build on can be classified into three categories: shape segmentation (Sec. \ref{sec:star_segmentation}), bounding boxes (Sec. \ref{sec:star_bboxes}), object packing (Sec. \ref{sec:star_packing}).

\subsection{Shape Segmentation}
\label{sec:star_segmentation}
3D object segmentation is fundamental for a number of applications, including shape analysis, modeling and recognition.
The choice of a strategy to subdivide an object into \emph{useful} parts, however, strongly depends on the context where these parts are expected to be used.
At the highest level, we differentiate between geometry-based and semantics-based segmentation.
From a geometric point of view, parts may be required to have well-defined characteristics with respect to some geometric property such as, for example, size, curvature or distance to a fitting plane. This approach to mesh segmentation is often used for texture mapping \cite{zhang2005}, remeshing \cite{sander2003}, simplification \cite{cohens2004}. In the context of 3D printing an important geometric requirement of each part is its size, and algorithms exist to partition the object so that each part is small enough to fit in the printing volume \cite{luo2012}. In a very recent work, Vanek and colleagues tackle the problem of reducing both printing time and support structures by segmenting an object and packing the resulting parts in a smaller volume \cite{vanek2014}.
Semantics-based segmentation is aimed at identifying parts that correspond to relevant features of the shape \cite{mortara2006}.  
Herewith we look for parts whose shape is likely to facilitate their packing and, to the best of our knowledge, no existing segmentation algorithm was designed with this objective in mind.

From another perspective, we may differentiate between single-resolution and hierarchical segmentation.
Single-resolution segmentations are appropriate either when segments are identified based on some geometric threshold (e.g. flatness), or when the number of segments is known in advance. Since in our setting we need to switch among different resolutions, we focus on hierarchical segmentation algorithms. The algorithm of \cite{katz2003} computes the so-called \emph{centrality} at each mesh vertex, and uses fuzzy clustering to recursively split the surface into smaller and smaller regions; for natural shapes such as humans and animals, the eventual parts typically correspond to functional areas of the object. Similar results were also obtained in \cite{reuter2009} by using Laplace-Beltrami eigenfunctions.
When man-made meshes need to be segmented, the bottom-up algorithm of \cite{attene2006} grows clusters starting from the single triangles driven by approximating geometric primitives such as planes, spheres and cylinders. A similar approach is employed in \cite{attene2008} when the object is represented by a tetrahedral mesh and the clusters are nearly-convex sets of tetrahedra.

For comprehensive overviews of mesh segmentation algorithms we point the reader to \cite{shamir2008} and \cite{chen2009}.

\subsection{Bounding boxes and box trees}
\label{sec:star_bboxes}
The calculation of bounding boxes is a fundamental problem in computer graphics, and can be formulated in several ways according to the bounding \emph{tightness} required. Axis-aligned bounding boxes (AABBs) can be computed very efficiently but are not sufficiently tight for many applications. Principal component analysis can be used to compute oriented bounding boxes (OBBs) that, though being tighter than AABBs, are still not optimal. Optimality is reached by minimum volume bounding boxes (MBBs) that, unfortunately, are rather difficult to compute. Currently, the best known algorithm is due to O'Rourke \shortcite{orourke1985} but its high computational complexity ($O(n^3)$ for $n$ input points) makes it virtually unusable for most practical problems.

Nonetheless, MBBs are so useful that even approximations are acceptable for some applications. In \cite{barequet1996} the shape is converted to a so-called \emph{box-tree} by recursively splitting it into parts bounded by approximate MBBs for the sake of collision detection. In this algorithm each approximate MBB is basically a refinement of the OBB, but no guarantees are given on the accuracy.
An algorithm to compute approximated MBBs with guaranteed accuracy is given in \cite{barequet2001}. This algorithm requires $O(n log n + n/\epsilon^3)$ operations with a relative accuracy of $\epsilon$.

\subsection{Packing}
\label{sec:star_packing}
Our problem is a variation of the bin packing problem \cite{ullman1973}; instead of minimizing the number of bins, we wish to minimize the volume of a single bin containing a set of appropriately placed objects. Some authors refer to this specific instance as to the Cutting and Packing problem \cite{chernov2010} or the Strip Packing Problem \cite{bansal2013}. Since these problems are known to be NP-complete \cite{johnson1974}, heuristic algorithms exist to find approximated solutions.

In two dimensions heuristics-based algorithms produce satisfactory results, and the underlying theory is well understood and exploited \cite{bennell2008}.
These studies were strongly motivated by industrial applications, including garment manufacturing and sheet metal cutting where the objective is to minimize the wasted material. In Computer Graphics, 2D packing algorithms are used for applications such as texture atlas generation \cite{levy2002} and for the construction of multi-chart geometry images \cite{sander2003}. In these cases both the container and the objects to be packed are raster images and the objective is to \emph{draw} all of them into a single small image.

In three dimensions, the packing problem arises in several industrial processes and the most noticeable algorithms are dedicated to rather specific cases in a purely practical perspective.
In \cite{egeblad2010}, for example, an algorithm is proposed to select several instances from a set of few furniture objects so that a given container can be filled with maximum efficiency. Though the packing efficiency obtained is very high (91.3\%), this approach is not exploitable in our setting because we need all the parts to be included and cannot assume to have several instances of a same part.
A more suitable formulation is given in \cite{wu2010} and \cite{bansal2013}, where all the 3D cartons of a given set are packed into a container of fixed base area and minimum height. Unfortunately these algorithms are based on the assumption that objects being packed are rectangular boxes, which simplifies the problem significantly.
As an exception, the 3D packing problem is tackled from a theoretical point of view in \cite{chernov2010}, where the concept of no-fit polygon is extended to the three dimensional setting. This approach, however, requires the input to be converted to so-called \emph{phi-objects}; though these objects efficiently represent geometric primitives such as cubes, spheres and cones, the representation of generic nonconvex 3D shapes may easily become so cumbersome to be computationally unmanageable.

\section{The split-and-pack approach}
\label{sec:splitpack}
Besides the object to disassemble, the input to our process includes a maximum number of parts and a target packing efficiency. Based on these, our algorithm calculates the parts and rotates and translates them so that they fit an axis-aligned box of mimimum volume. Notice that, since the object volume is constant, maximizing the packing efficiency is equivalent to minimizing the box volume. Thus, the lower bound for the target packing efficiency can be actually converted to an upper bound for the eventual box volume.

In principle, the calculation of the parts might exploit existing segmentation algorithms (e.g. \cite{luo2012}), but in most cases a single fixed sementation leads to a sub-optimal packing. We have verified that searching the best segmentation within an appropriate binary hierarchy provides much tighter packings (see Sec. \ref{sec:results} and Fig. \ref{fig:choppercomp}). Instead of relying on a hierarchy, an alternative approach is used in \cite{vanek2014} where parts in an over-segmentation are merged in different configurations to find the most effective packing.

\begin{figure}
   \includegraphics[width=\linewidth]{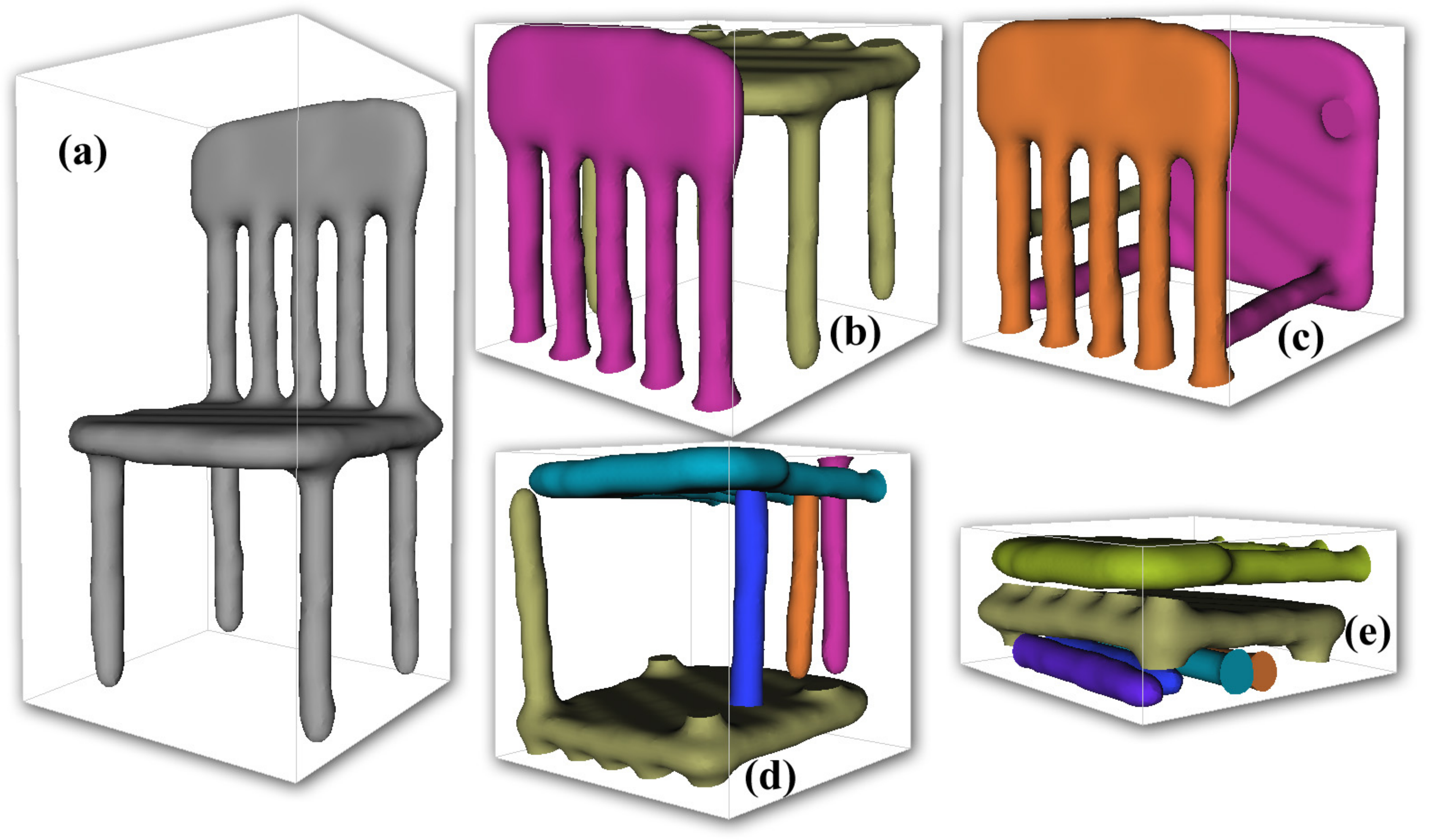}
   \caption{The split-and-pack approach. According to the target efficiency, the whole chair model could not be packed without splits (a). Thus, the model was split in two parts (b), then in three (c), four (d), and five parts. When split in six parts (e) the model could be packed with the required efficiency.}
   \label{fig:chairs}
\end{figure}

Having said that, our strategy to determine the best number of parts is based on the calculation of a hierarchy of parts organized into a binary tree. We start by analyzing the root of the tree (i.e. the whole object): if it can be packed into its minimum volume bounding box with a sufficient efficiency (according to the target), then we have a solution and stop. Otherwise, if the wasted space is too much we replace the root with its two children (i.e. two parts that build the entire object), compute their best packing and the corresponding efficiency. If it is sufficient the process stops, otherwise we go down further in the hierarchy and replace one of the two nodes with its two children, compute the best packing of the three parts and so on (see Fig. \ref{fig:chairs}).
The process may stop either when the packing efficiency reaches the predetermined target or when the number of parts exceeds the maximum. In this latter case the user is warned that no solution can be found, and is allowed to increase the maximum number of parts.
Note that when the algorithm replaces a node with its two children, the adjacency area of the corresponding two parts is refined to guarantee that parts can be eventually reassembled without obstructions (see Fig. \ref{fig:planefit}).

One may argue that the number of parts could have been the unique parameter to drive the global minimization of the box volume. In principle, indeed, the packing efficiency should not decrease as the number of parts grows. In practice, however, object packing is an NP-hard problem and our algorithm necessarily relies on heuristics that may easily lead to violations of this rule. Furthermore, if the object shape is already very close to a parallelepiped (according to the target) we prefer not to split it.
On the other hand, we could have used the target efficiency as the unique parameter, but such a choice makes the algorithm very unpractical when a large number of parts is necessary to reach the target. Computing the packing as described in Sec. \ref{sec:packing}, indeed, is a quite expensive operation. Hence, in these cases we prefer to let the user choose whether to increase the number of parts or to be happy with the lower packing efficiency obtained so far.

To summarize, the overall split-and-pack process can be described by algorithm \ref{al:splitandpack}.

\begin{algorithm}[t]

\caption{The overall split-and-pack algorithm}
\label{al:splitandpack}
\begin{algorithmic}[1]

\REQUIRE A 3D object $S$, a maximum number of parts $N_{max}$, and a target packing efficiency $E$.
\ENSURE A set of roto-translated parts $S_i$ and the corresponding axis-aligned bounding box $B$.

\STATE $Vol_{Max} := Vol(S)/E$
\STATE Create a hierarchical box-based segmentation $H$ of $S$ (see Sec. \ref{sec:segmentation})
\STATE $N_S := 1$
\STATE $S_0 := S$
\STATE $B := MBB( S )$
\WHILE {$Vol(B) > Vol_{Max}$}
	\IF {$N_S = N_{Max}$}
		\STATE Warn and ask user a new value for $N_{Max}$
		\IF {no new value provided}
			\STATE terminate
		\ENDIF
	\ENDIF
	\STATE Split one of the $S_i$ according to $H$ and increase $N_S$
	\STATE Refine the adjacency area (Sec. \ref{sec:reassembling})
	\STATE Compute best packing and corresponding $B$ (see Sec. \ref{sec:packing})
\ENDWHILE
\STATE Perforate thin parts (Sec. \ref{sec:reassembling})
\end{algorithmic}
\end{algorithm}

\section{Box-based hierarchical segmentation} \label{sec:segmentation}
Our approach to generate a hierarchy of \emph{box-like} parts out of a given object relies on the assumption that the object is actually a solid. This means that surface meshes with holes, self-intersections and other sorts of \emph{defects} must be processed in advance using appropriate mesh repairing tools \cite{repsurvey}. When the input mesh is guaranteed to bound a solid, it is converted to a tetrahedral mesh by constrained Delaunay triangulation \cite{si2005} and its tetrahedra are hierarchically clustered using a bottom-up approach inspired on \cite{attene2008}.

Specifically, let $M$ be the tetrahedral mesh and let $T$ be the set of its tetrahedra. The dual graph $D=(N,A)$ of $M$ is defined as follows: each node in $N$ corresponds to a tetrahedron in $T$, and there is an arc (dual edge) in $A$ connecting two nodes in $N$ if the corresponding tetrahedra in $M$ share a triangular facet. Now, if one considers each node of such a dual graph to represent a cluster of connected tetrahedra (initially made of a single element), merging two tetrahedra into a single representative cluster corresponds to contracting a dual edge into a single node, that is, the two nodes of the arc are identified and the adjacency relations are updated accordingly in the graph data structure.

In our algorithm, a priority queue is created in which all the dual edges are sorted based on the \emph{cost} of their contraction. At each step, the dual edge with lowest cost is popped from the queue, it is contracted, and all the edges incident to the new representative node are updated, that is, their cost is re-computed and their position in the queue is updated accordingly.
The method produces a hierarchy which can be represented by a binary tree of clusters where the root identifies the whole tetrahedrization, and the leaves are the individual tetrahedra.

\begin{algorithm}[t]

\caption{Hierarchical segmentation based on absolute aboxiness}
\label{al:htc}
\begin{algorithmic}[1]

\REQUIRE A tetrahedral mesh $T$
\ENSURE A binary tree of tetrahedra clusters

\STATE Create a dual vertex and a corresponding singleton cluster for each tetrahedron in $T$
\STATE Create a dual edge for each pair of tetrahedra sharing a facet
\STATE Associate a cost to each dual edge (see Sec. \ref{sec:boxiness})
\STATE Create a heap of all the dual edges sorted based on their associated cost
\WHILE {heap is not empty}
	\STATE pop the first edge from the heap (let it be $e$)
	\STATE Contract $e$ to a single dual vertex $v$ and merge the corresponding clusters
	\STATE Update the cost of all the edges incident to $v$ and update their position in the heap
\ENDWHILE

\end{algorithmic}
\end{algorithm}

\subsection{Boxiness}
\label{sec:boxiness}
Clearly, the just-exposed algorithm may create different hierarchies as the \emph{cost} associated to dual edges changes.
We remind that our final objective is to fit the parts into a container while minimizing the wasted space. Since equal rectangular cuboids completely fill the space, they are good candidate shapes to look at. From now on the shorter term \emph{box} is used to refer to a rectangular cuboid, whereas the similarity of an arbitrary polyhedron with a box is called \emph{boxiness}.
A natural way to define the boxiness $B(P)$ of a polyhedron $P$ is as follows:

\begin{equation}
\label{eq:boxiness}
B(P) = Vol(P)/Vol(MBB(P))
\end{equation}

where $Vol(.)$ denotes the volume and $MBB(.)$ denotes the minimum volume bounding box \cite{orourke1985}.
According to definition \ref{eq:boxiness} the boxiness is $1$ only if $P$ is an actual box, while it is smaller in all the other cases.

In a first experiment we tried to use the concept of boxiness directly to define the cost to be associated to dual edges in our hierarchical clustering. Specifically, the cost of an edge $e=\{N(C_1), N(C_2)\}$ was defined as $1-B(C_1 \cup C_2)$, where $N(C_i)$ is the node in $D$ corresponding to cluster $C_i$. Unfortunately this approach easily leads to extremely unbalanced binary trees that, besides making the algorithm exceptionally slow, are not really practical and useful in our setting.
So, we have chosen an alternative definition of \emph{absolute aboxiness} $A(P)$ as follows:

\begin{equation}
\label{eq:aboxiness}
A(P) = Vol(MBB(P))-Vol(P)
\end{equation}

\begin{figure}
\centering
   \includegraphics[width=0.8\linewidth]{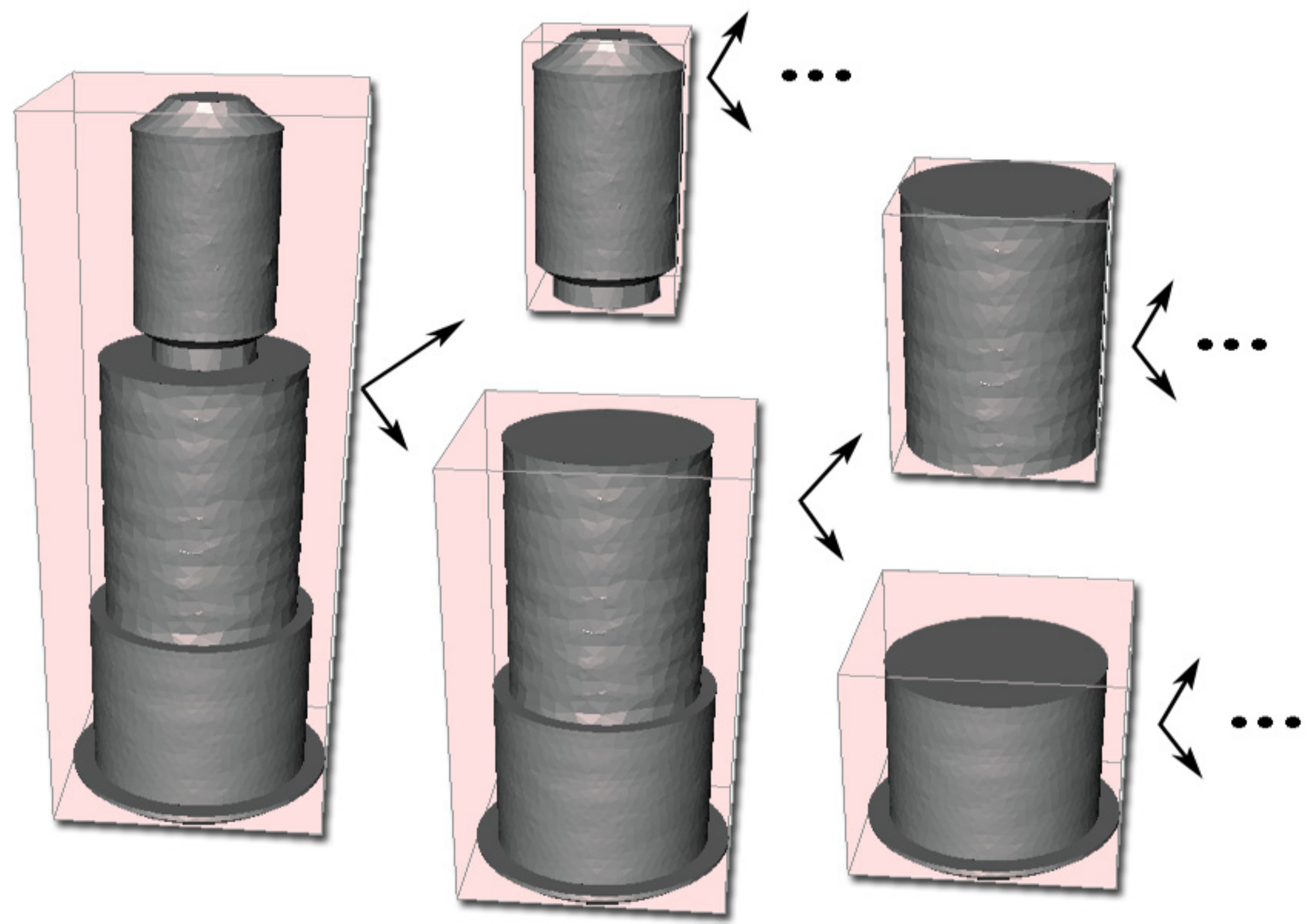}
   \caption{A tetrahedral mesh hierarchically segmented using our approach. Though the algorithm proceeds bottom-up by clustering tetrahedra, the most interesting parts of the resulting tree are close to the root, and the tree itself can be interpreted as a sequence of splits. The whole mesh (left) is split in two submeshes (middle), and each of them is further split in two other submeshes (right), and so on, down to the single terahedra. Each part in the hierarchy is depicted along with its corresponding minimum bounding box.}
   \label{fig:htc}
\end{figure}

According to this new definition, the absolute aboxiness is $0$ only if $P$ is an actual box, while it is larger in all the other cases.
In a second experiment we defined the cost of an edge $e=\{N(C_1), N(C_2)\}$ as $A(C_1 \cup C_2)$ with the twofold advantage of producing balanced trees and improving the algorithm performances, therefore this has been our final choice.
An example binary tree resulting from this procedure is partially depicted in Fig. \ref{fig:htc}.
Note that if $P$ is a hollow box, then $A(P)$ is greater than zero, and this is desirable because we wish to split such objects to pack them more efficiently.

\section{Shape packing} \label{sec:packing}
When a set of parts is available, our objective is to determine a roto-translation for each part so that their overall minimum-volume bounding box is minimized. Notice that such a minimum bounding box can, in its turn, be rotated along with its inner parts to make it axis-aligned. Thus, without loss of generality, our problem is to determine roto-translations of the parts that minimize their overall axis-aligned bounding box.

Our approach to solve this problem starts from an initial axis-aligned box that can contain the entire unpartitioned object. Prior to computing such a box, however, the object is rotated according to its minimum-volume bounding box, so that the initial box is both axis-aligned and of minimum volume. Then, we consider one of the three axes (e.g. the $Z$, if this is the direction for which the box is mostly extended) to be the \emph{vertical} direction and define the two box faces which are orthogonal to this direction as the lower and upper bases (e.g. the faces with minimum and maximum $Z$ respectively).

\begin{figure}
   \includegraphics[width=\linewidth]{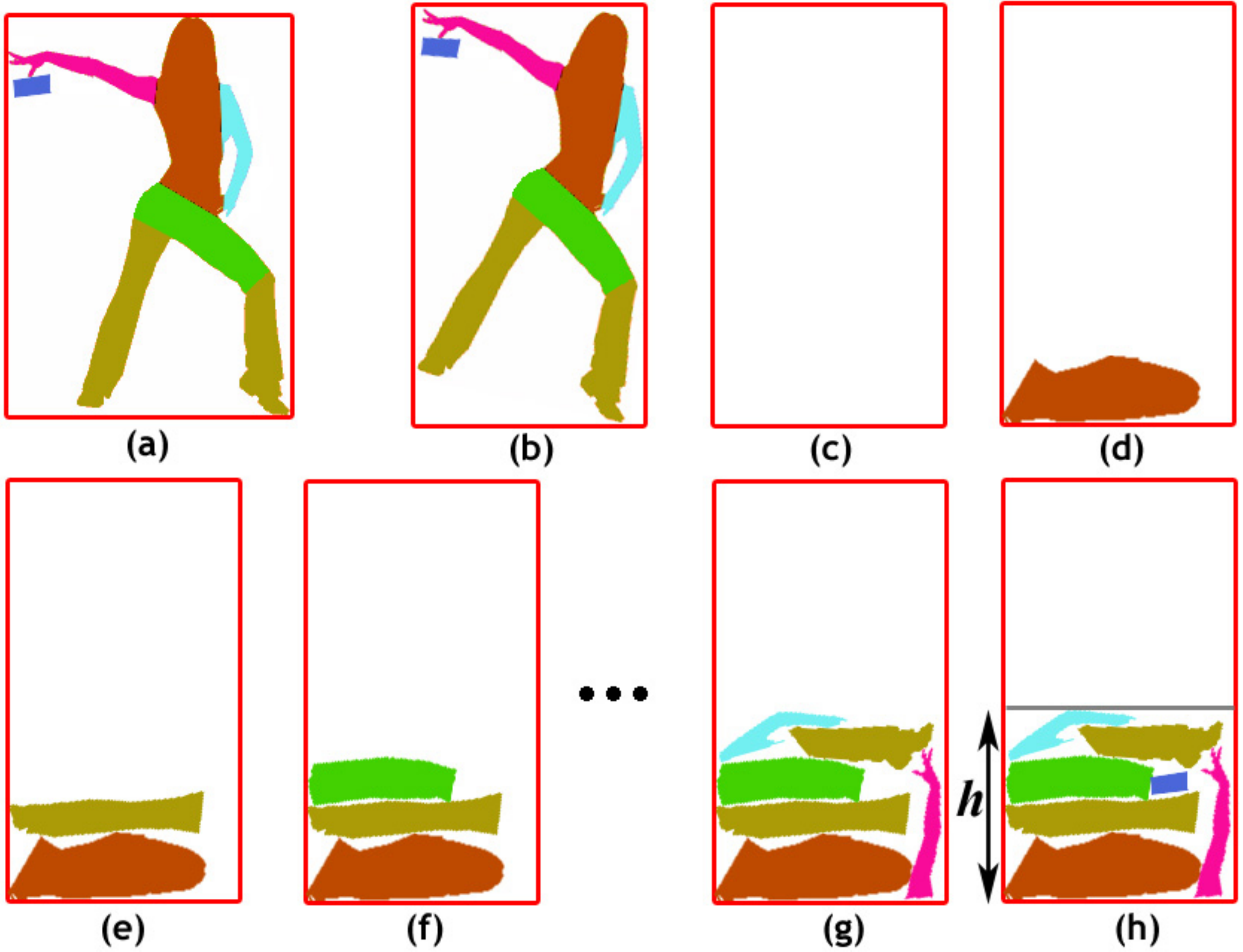}
   \caption{An input 2D shape (a) is rotated so that its AABB corresponds to its MBB (b). The AABB is used as the initial empty container (c), and parts are inserted one by one from largest to smallest (d)-(g). In this example only the last part can be placed into a hole (h), whereas all the others are placed on top of previously inserted parts.}
   \label{fig:packing2d}
\end{figure}

Then, parts are inserted one by one while measuring their total height $h$, which is the size of their current axis-aligned bounding box along the vertical direction (Fig. \ref{fig:packing2d}). When a part is inserted, the objective is to minimize the overall height. Thus, the first part is rotated so that its vertical extension is minimized. After such a rotation, the first part is pushed toward the lower-left corner of the box lower base. Clearly, all the parts must be entirely contained into the box, hence roto-translations that violate this rule are not considered. The placement of each subsequent part is slightly more complex because it must consider the previously occupied space, but the objective of minimizing the overall total height remains the same.
We note that in most cases the insertion of a new part does not increase the height at all, thus we must select the most suitable roto-translation based on other criteria. With the objective of minimizing the wasted space, we look for appropriate portions of unoccupied space to host the part. Specifically, if previously placed parts leave free \emph{holes} among them, we identify the hole that can best contain the next part, meaning that the difference between the hole's and part's volumes must be minimized. If no such hole exists (i.e. there are no holes or none of them can host the part), the new part is placed on top of the others. This procedure is described in more depth in Sec. \ref{sec:singleplacement} and illustrated in Fig. \ref{fig:packing2d}.

We observe that part insertion order influences the eventual packing efficiency (Table \ref{tab:results}). From a theoretical point of view, indeed, for random insertion orders we can expect a difference with respect to the optimal solution of up to 70\% (\cite{ullman1973}). Conversely, if parts are inserted from largest to smallest such a difference decreases down to 22\% (\cite{johnson1974}). Thus, the first step in our packing algorithm is a sorting of the parts based on their maximum extension which, for each part, corresponds to the maximum length of its minimum volume bounding box.

\subsection{Placement of a single part}
\label{sec:singleplacement}
Inserting a part $P$ into the box amounts at determining a roto-translation of $P$ which, in our setting, is defined by a quaternion $Q$ and a vector $V$ for the rotational and translational parts respectively. Operationally, we first roto-translate $P$ so that its minimum volume bounding box becomes axis-aligned and centered at the origin. Then, we rotate $P$ by $Q$ and translate it by $V$ in this order.

Determining the optimal $Q$ and $V$ is rather complex thus, for the sake of simplicity, we first outline our approach to compute $V$ for a given fixed value of $Q$.

Our algorithm proceeds as follows:
\begin{itemize}
\item If there are holes that can contain $P$, we select the one whose volume is closer to $P$'s volume (see Sec. \ref{sec:holes}) and set $V$ accordingly.
\item If no such hole exists, we set $V$ so that $P$'s underlying free volume is minimized (see Sec. \ref{sec:frontier}) while not increasing $h$ .
\item If all the positions increase $h$, we set $V$ so that $h$ increases as few as possible. When several positions produce the same minimum increase of $h$, we select the one that minimizes $P$'s underlying free volume (see Sec. \ref{sec:frontier})
\end{itemize}

In practice, to compute the second and third points in the afore-mentioned procedure we use a uniform discretization (i.e. a voxelization) of the box. Besides making the problem algorithmically tractable, the involved height and volumes become integer numbers with upper bounds, and this allows us to define a single \emph{cost} to be minimized as follows:

\begin{equation}
\label{eq:position_cost}
Cost(P) = \Delta_h B + U
\end{equation}

where $\Delta_h$ is the increase of $h$, $U$ is $P$'s underlying free volume, and $B$ is the total volume of the box.

To include $Q$ in the cost minimization process, we simply add an outer loop where a set of pre-defined rotations is iteratively assigned to $Q$. Pre-defined rotations are computed by uniformly sampling the space of unit quaternions, which is equivalent to uniformly sampling $\mathbb{S}^3$. To do this, we employ the following approach by Marsaglia (\cite{marsaglia72}): select $x_1$ and $y_1$ uniformly in $(-1,1)$ until $s_1 = x_1^2 + y_1^2 < 1$; similarly, select $x_2$ and $y_2$ uniformly in $(-1,1)$ until $s_2 = x_2^2 + y_2^2 < 1$; then

\begin{equation}
\label{eq:uniform_quat}
\mathbf{q} = [x_1, y_1, x_2 \sqrt{(1-s_1)/s_2}, y_2 \sqrt{(1-s_1)/s_2}]
\end{equation}

is uniformly distributed on $\mathbb{S}^3$.
Before applying any rotation each part is made \emph{axis-aligned}, which means that it is rotated so that its MBB becomes axis-aligned. This is expected to  maximize the possibilities of good fits within an axis-aligned container.

\subsubsection{Placement into holes}
\label{sec:holes}
At any stage of the algorithm, free voxels of the box can be clustered into regions of two types that we call \emph{holes} and \emph{slots}.
A free voxel belongs to a hole if there exists at least a non-free voxel over it. Conversely, a free voxel belongs to a slot if all the voxels over it (up to the upper base) are also free. Thus, a hole is a maximal connected set of free voxels having a ``ceiling'' of non-free voxels.

Let $P$ be the part to be placed, and let $V$ and $Q$ define its current position and orientation. In order to check if $V$ and $Q$ define a valid roto-translation for $P$ we simply check that it does not intersect any previously placed part and completely belongs to the box. To do that, we simulate a rasterization of $P$ into the voxel grid and verify that it involves only free voxels: in this case we say that $P$ is \emph{rasterizeable}. Note that we rasterize the actual shape of the part, and not its bounding box.

\begin{figure}
\centering
   \includegraphics[width=0.8\linewidth]{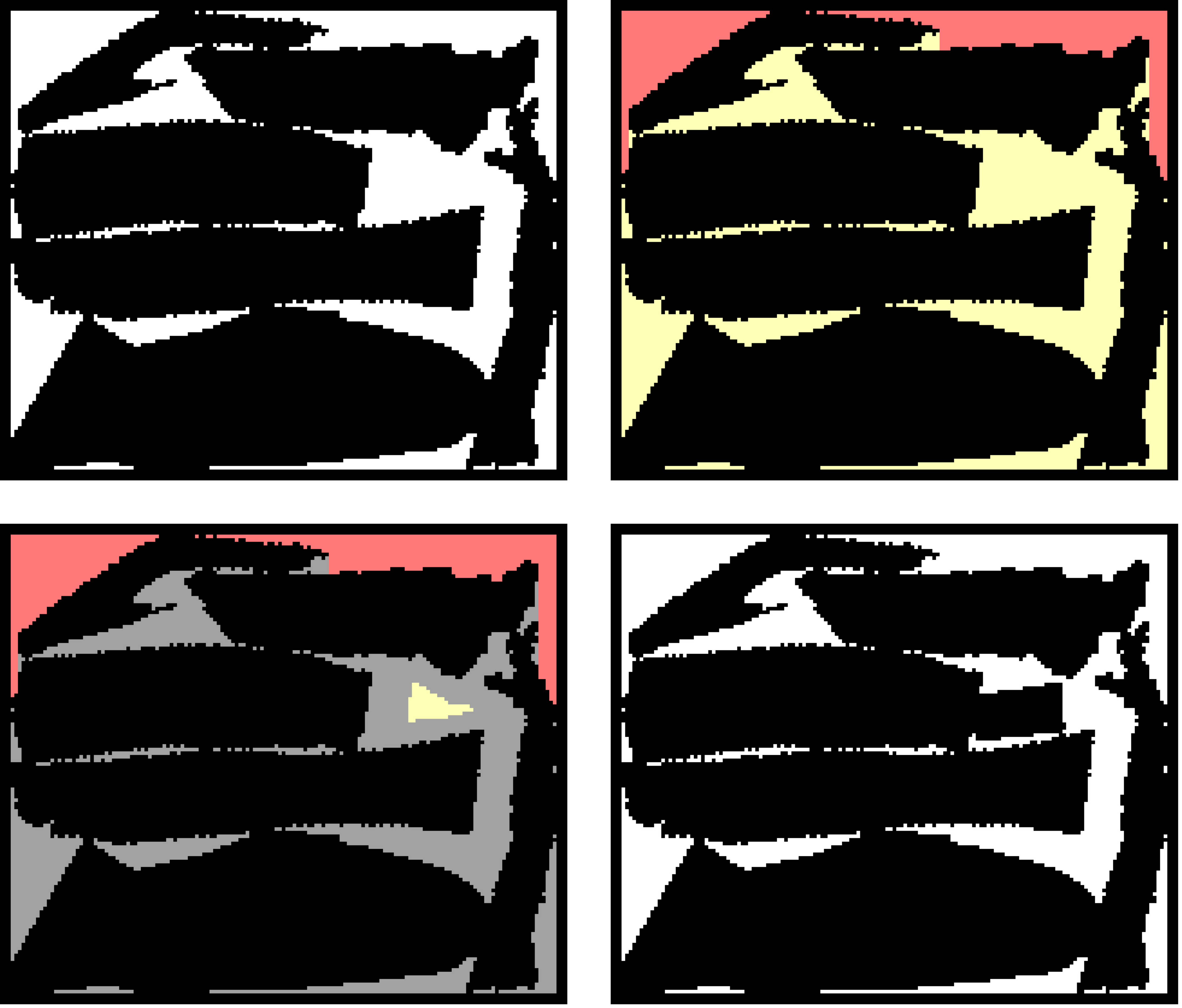}
   \caption{Detailed operations for the placement of the last part in Fig. \ref{fig:packing2d}. Based on the previous configuration (top-left), voxels are classified in holes and slots (top-right). Then, hole voxels are filtered based on the shape of the part to be inserted (bottom-left). Finally, the part is placed in the first unfiltered voxel around which the part is rasterizeable.}
   \label{fig:holes}
\end{figure}

When $P$ is given with a fixed orientation, we wish to see which $V$ among all the valid positions fills a hole while minimizing its remaining free volume. A naive approach considers all the voxels belonging to holes as potentially valid positions and looks for the best one. Notice that this method is not as bad as it might seem at a first sight because the virtual rasterization simply stops at the first occurrence of a non free voxel. However, we can significantly reduce the search space by skipping voxels which are too close to the border of the hole; specifically, we consider $P$ in its current orientation $Q$ as being centered at the origin, and see which is its farthest point on each of the six cartesian semi-axes. These six distances are then used to actually shrink all the holes (see Fig. \ref{fig:holes}) so that only a minority of voxels needs to be checked.

\subsubsection{Placement on top of other parts}
\label{sec:frontier}
When no suitable hole can be found, we look for a valid roto-translation that minimizes the underlying free volume $U$ of $P$.
This approach is inspired on \cite{sander2003} and allows to accurately wedge concave parts in the free space left by other parts.
Formally, $U$ can be defined as follows:

\begin{equation}
\label{eq:undfreevol}
U = \int_{\Gamma} (\lfloor P(\mathbf{x}) \rfloor - \lceil \mathbb{P}(\mathbf{x}) \rceil) \partial \mathbf{x}
\end{equation}

where $\Gamma$ is the $2D$ projection of $P$ onto the box lower base, $\lfloor P(\mathbf{x}) \rfloor$ is the lowest point of intersection of $P$ and the vertical line defined by $\mathbf{x} = <x,y>$, $\mathbb{P}$ is the union of all the previously placed parts, and $\lceil \mathbb{P}(\mathbf{x}) \rceil$ is the upper point of intersection of $\mathbb{P}$ and the vertical line defined by $\mathbf{x} = <x,y>$.

In our discrete setting, the integral in Equation \ref{eq:undfreevol} becomes a sum, and $\mathbf{x}$ ranges within the discretized version of $\Gamma$.
The function $\lceil \mathbb{P}(\mathbf{x}) \rceil$ is initially zero everywhere and it is updated after the placement of each single part.

To determine the roto-translation that minimizes $U$ we simply iterate among a sampling of the rotations and, in a nested loop, we iterate again among the set of positions that make $P$ tangent with $\lceil \mathbb{P}(\mathbf{x}) \rceil$.

\subsection{Overall packing algorithm}
To summarize, the initial box $B$ is set so that it can contain the entire unsplit object, whereas its $height$ is minimized through algorithm \ref{al:packing}.
Additionally, we notice that the inserted parts do not necessarily fill the entire box extension along the $X$ and $Y$ directions; in this case the $X$ and $Y$ box sizes are reduced down to actual tangency and the efficiency is re-computed.

In principle this procedure can be used to perform the packing, but since the lower-base is fixed it may easily lead to too coarse results. Thus, we add an additional outer loop to this algorithm that iterates over various shapes of the lower base and keep the best result. Specifically, we consider four growing factors of 0\% (=no grow), 25\%, 50\%, -25\% (=reduction factor), and apply them to both side $X$ and side $Y$ of the lower base in all the $16$ possibile combinations; for each of these transformations the box initial height is increased so that its volume remains constant. The number of growing factors can be increased by the user to strive for better efficiency, though in our experience 16 combinations are a good tradeoff between accuracy and computing time.
In this phase possible additional constraints regarding the base size can be taken into account (e.g. application-dependent upper bounds for one or both its dimensions).

\begin{algorithm}[t]

\caption{The height-minimizing algorithm. MBB(.) denotes the minimum volume bounding box, whereas $R$ is a sampling of the rotation space, and $R_q(P)$ is the result of rotation $q$ applied to the axis-aligned version of $P$. $P$ is said to be rasterizeable if its rasterization does not intersect previously-occupied voxels.}
\label{al:packing}
\begin{algorithmic}[1]

\REQUIRE A set of $n$ parts $P_i$ constituting an assembly $T$
\ENSURE A set of roto-translations $R_i$ to be applied to the $P_i$

\STATE Compute a minimum axis-aligned container $B$ for the overall $T$
\STATE Create a raster 3D image $I$ of $B$
\STATE Create a list $H$ of voxels classified as $holes$ (initially empty)
\STATE Translate each of the $P_i$s so that the center of its MBB coincides with the origin
\STATE Create a sorted stack $L$ of all the $P_i$s
\STATE Initialize the 2D image $\lceil \mathbb{P}(\mathbf{x}) \rceil$ to zero
\WHILE {$L$ is not empty}
	\STATE pop the first part from $L$ (let it be $P$)
	\FORALL {$q \in R$}
		\STATE Create a restricted version $H_q$ of $H$ according to $R(P_i)$
		\FORALL {$v \in H_q$}
			\IF {$R(P_i)$ is rasterizeable in position $v$}
				\STATE Compute the hole wasted space
			\ENDIF
		\ENDFOR
	\ENDFOR
	\IF {One or more holes could be found}
		\STATE Actually roto-translate $P_i$ using the $q$ and $v$ that minimized the wasted space
	\ELSE
		\FORALL {$q \in R$}
			\FORALL {$v \in \lceil \mathbb{P}(\mathbf{x}) \rceil$}
				\STATE $\hat{v} := v+s$, where $s$ is the minimum vertical translation that makes $R(P_i)$ rasterizeable
				\IF {a valid $s$ exists}
					\STATE Compute the Cost according to equation \ref{eq:position_cost}
				\ENDIF
			\ENDFOR
		\ENDFOR
		\IF {One or more positions could be found}
			\STATE Actually roto-translate $P_i$ using the $q$ and $v$ that minimized the Cost
		\ELSE
			\STATE Terminate with failure (no arrangement could be found)
		\ENDIF
	\ENDIF
	\STATE Rasterize $P_i$ in $I$ and update $H$ and $\lceil \mathbb{P}(\mathbf{x}) \rceil$ accordingly
\ENDWHILE

\end{algorithmic}
\end{algorithm}

\section{Printing and reassembling the parts} \label{sec:reassembling}
\begin{figure}
   \includegraphics[width=\linewidth]{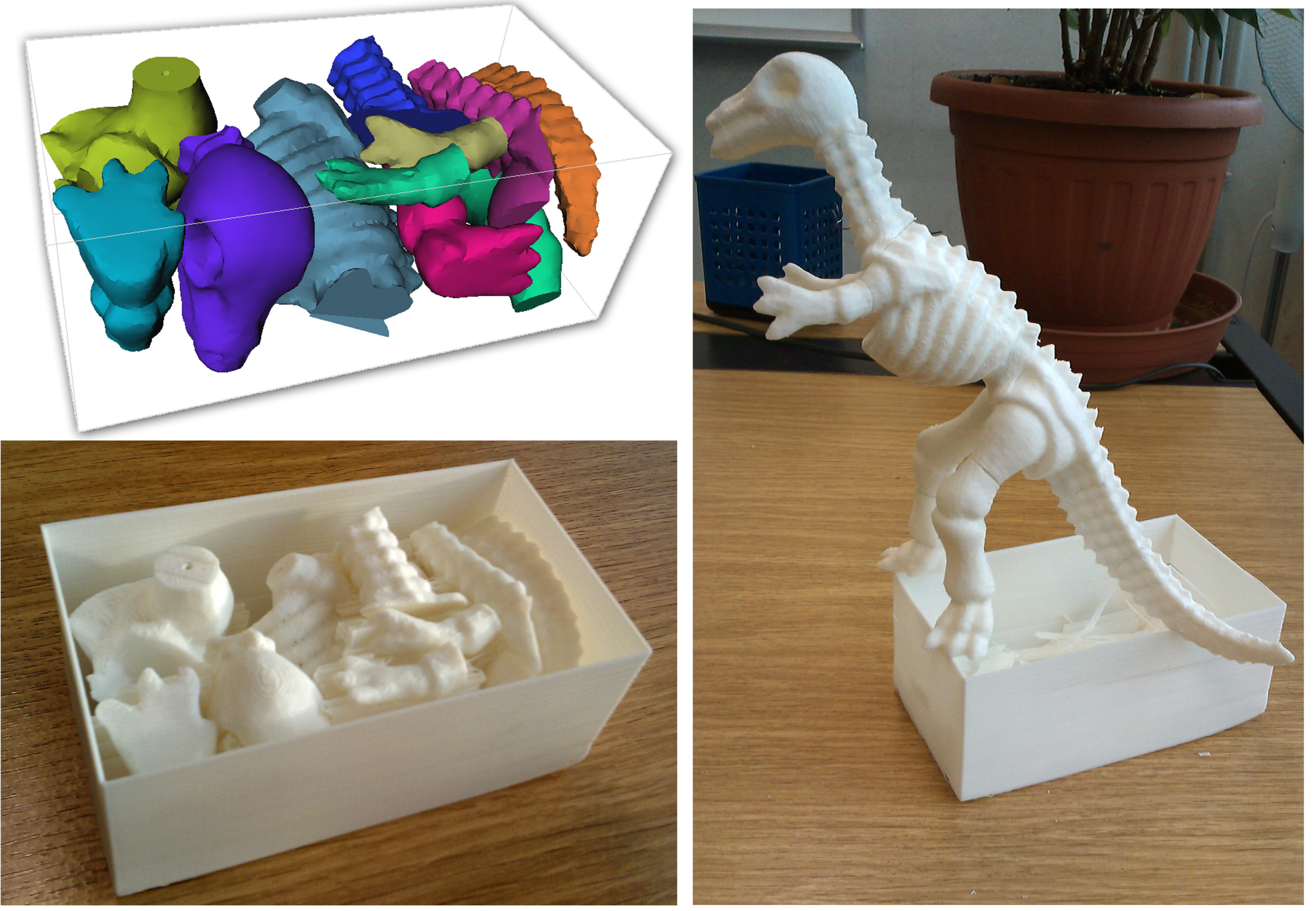}
   \caption{Parts computed and packed by our algorithm (top-left), the printed box with parts inside (bottom-left), and the reassembled object on its empty box (right). Parts could be manually extracted from the box and cleared of all the support structures in less than three minutes.}
   \label{fig:dino}
\end{figure}

Once the parts and their arrangement are available, the process to generate the physical prototype depends on the approach employed by the printer. For printers that use so-called \emph{powder beds} (e.g. Z Corporation products) the parts can be printed alltogether at once in their packed position, and the box itself can be printed around the parts without the top (Fig. \ref{fig:cessna}). Such a printed box can then be removed from the printing chamber with all its contents, packed, and possibly shipped without the need to remove the unbound powder that serves to keep the parts from moving around during transportation. At destination, the parts can be removed from the box along with the extra powder, while taking particular care if tiny parts are contained.
Printers based on a \emph{fused-deposition model} (e.g. the widely diffused products by MakerBot) need to insert additional material to support parts that otherwise would fall down during printing. These support structures are very thin plastic sheets, and we have verified that in most cases they can be simply teared off with fingers (e.g. Fig. \ref{fig:dino}). In a few more difficult cases scissors and box cutters were necessary to avoid breaking the more delicate object parts, such as Neptune's fingers in Fig. \ref{fig:teaser}.
In our packed boxes, support structures are attached to nearly the whole surface of the parts, with the exception of the upper surface over which no other part is placed.
To minimize the manual work required to remove support structures, before printing we reorient the box so that its vertical extension is the minimum among its width, height and depth.
This is equivalent to maximize the lower base area, and thus is expected to maximize the total upper surface. Furthermore, such an orientation better fits the printing volume of the most diffused printers, and thus allows to print larger boxes.
Herewith we assume that the box fits the printing volume: if it does not, parts can still be printed one by one and then packed as described by the algorithm.

\begin{figure}
   \includegraphics[width=\linewidth]{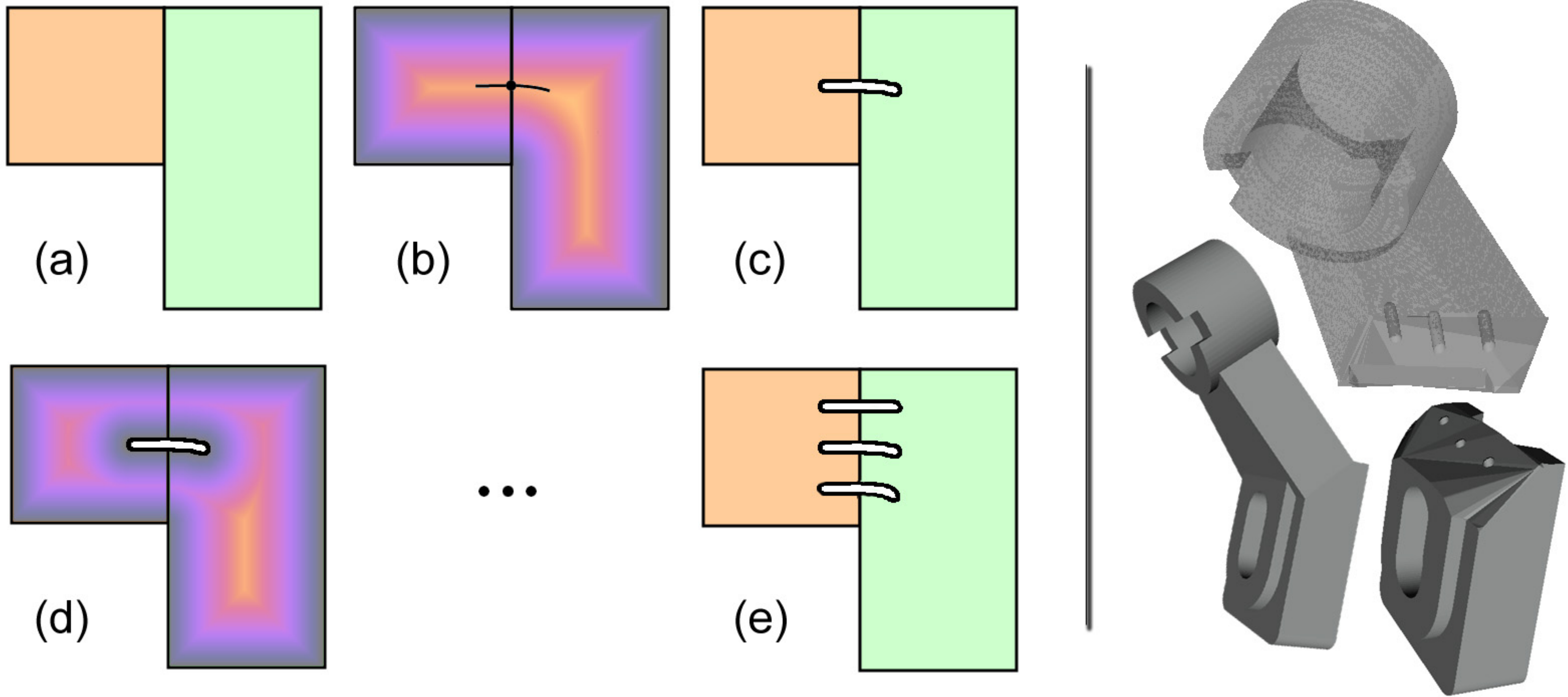}
   \caption{A 2D schematic example of the perforation algorithm (left) and an actual 3D example (right).}
   \label{fig:perforation}
\end{figure}

To reassemble the parts one may choose among various approaches, including printed connectors \cite{luo2012}, glue or screws. Though connectors are quite practical in principle, there are configurations (e.g. thin object parts) for which no useful connector can be constructed. Furthermore, in analogous cases glue does not guarantee sufficient robustness if used alone. Thus, our proposal is to use a combination of glue and iron wire, so that even parts as thin as the iron wire diameter (plus a small tolerance) can be effectively assembled. 
To do that we need to virtually \emph{perforate} the parts before printing them, and the parameters to be set are the diameter $d_w$ and length $l_w$ of the iron wire pieces to be used at reassembly time. Note that since iron wire is relatively flexible, the perforations need not to be perfectly straight.
To compute each perforation, we consider the two adjacent parts that are to be connected in their original position (Fig. \ref{fig:perforation} (a)), compute their overall signed distance function $D$ and select a point $P_0$ on their adjacency area for which $D$ is maximum. Starting from $P_0$, we orthogonally leave the adjacency area and track the straightest path that follows the maximum of $D$ (Fig. \ref{fig:perforation} (b)) up to an arc distance $l_w/2$ in both directions from $P_0$ \cite{cornea2007}. If the path passes through a point for which $D$ is smaller than $d_w$, it means that a smaller wire must be used to fit the part thickness. Otherwise the path is accepted and we simply perform a boolean difference \cite{bernstein2009} between the original part and a $(d_w/2)$-offset of the path (Fig. \ref{fig:perforation} (c)).
After such a first modification, we update $D$ (Fig. \ref{fig:perforation} (d)) and repeat the process on the same adjacency area to create a new perforation. Such an iterative process stops either when the distance between the perforation and the object surface becomes smaller than a given threshold, or when a maximum number of perforations is performed, whichever occurs first (Fig. \ref{fig:perforation} (e)). A practical example on a 3D object is shown on the right in Fig. \ref{fig:perforation}.

\begin{figure}
   \includegraphics[width=\linewidth]{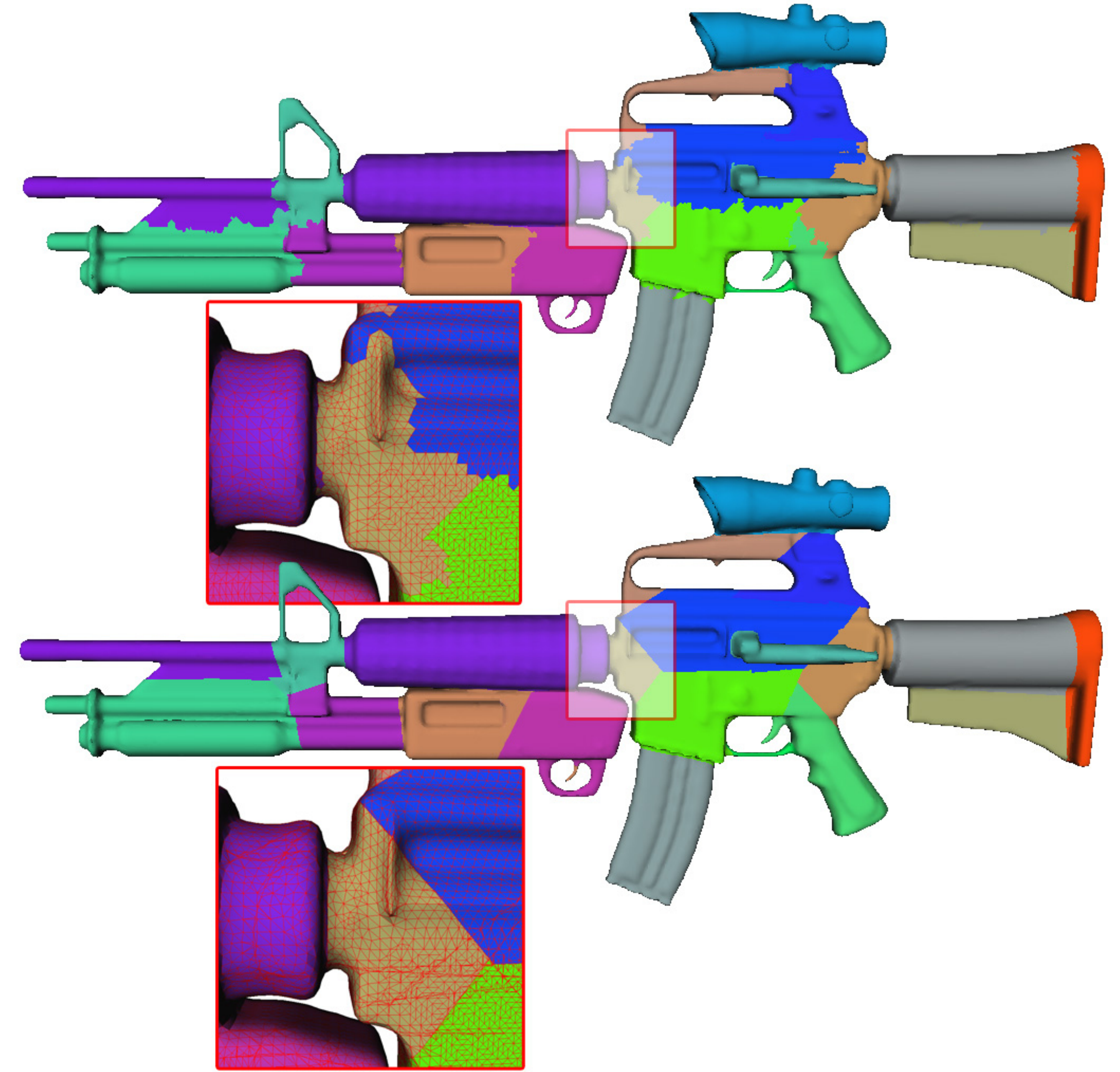}
   \caption{Parts produced by our segmentation algorithm might be pairwise unassemblable (top). We can detect and solve this problem by replacing the split areas with their best fitting planes (bottom).}
   \label{fig:planefit}
\end{figure}

Note that, since the split areas are not necessarily flat, it is worth analyzing them because corresponding part pairs may be eventually unassemblable. It is necessary that each split surface $M_i$ can be represented as a height field, which means that there exists a direction along which the two parts may be moved without touching each other until they match. To do such a test, we compute a ``discrete'' Gauss map as a finite set of points on the unit sphere $S^2$ representing the $M_i$'s facet normals. Then, we check whether the convex hull of these points contains the origin: if it does, for each possible direction $d$ at least one of the normals points in the opposite direction (i.e. its dot-product with $d$ is negative), which means that $M_i$ cannot be represented as a height field.
Thus, before running the packing algorithm, the segmentation is modified by cutting tetrahedra along $M_i$'s best-fitting plane. Actually such a cut is naturally close to the original $M_i$ due to our aboxiness minimization, therefore we simply perform the cut for all the adjacency surfaces with no need to analyze the Gauss map (see Fig. \ref{fig:planefit}). All the segmentations depicted in this paper are refined as described, with the exception of Fig. \ref{fig:perforation}-right which shows that the perforation algorithm may work even on nonflat areas.

To reassemble the object in the right order, one can simply traverse the binary tree of the parts from bottom to top. Thus, the first two parts to be merged are the last two leaves that, after merge, form an intermediate node of the tree that can then be merged with another node, up to the root of the tree. In this way, the tree can be used to automatically generate assembly instructions.

\section{Results and discussion}
\label{sec:results}
\begin{figure*}
   \includegraphics[width=\linewidth]{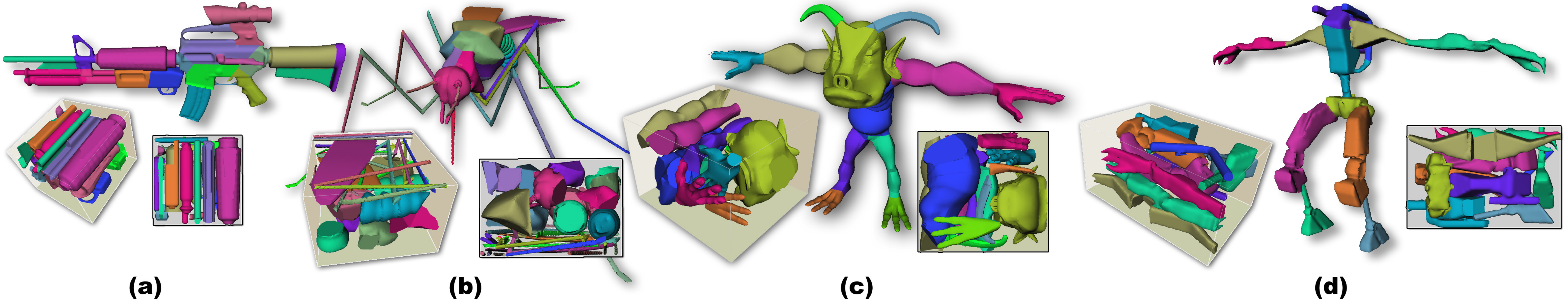}
   \caption{Some example models used to test our algorithm. The resulting part packing is shown both in perspective and in orthographic view to better convey the packing efficiency.}
   \label{fig:compound}
\end{figure*}

To run our experiments, we have implemented a prototype C++ program where Barequet's approach \cite{barequet2001} is employed to compute the minimum volume bounding boxes. Equation \ref{eq:position_cost} required a careful implementation to avoid overflows: when the box is a $256^3$ voxel cube, the cost may grow over $2^{32}$ and a 64-bit integer is required. Experiments were run on a MS Windows 7 OS, 1.87 GHz Intel Core 2 PC with 4 Gb RAM, the rotation space was sampled using 10 uniformly-spaced rotations, and the initial bounding box is discretized using $256^3$ voxels. These parameters provide a good tradeoff between time and efficacy (our slowest test took < 6 minutes, including the 56 seconds used to produce the hierarchy). The largest model in our test set is the mosquito (Fig. \ref{fig:compound}(b), 216299 tets). The user can increase the default number of rotations and the number of \emph{growing factors} for the box base to possibly improve the results at the cost of a linearly growing computing time for the packing phase. The iron wire diameter $d_w$ and length $l_w$ were set to 1 mm and 12 mm respectively, and our implementation was set to produce a single perforation on an adjacency area only if the maximum value of the inner distance field $D$ is less than 10 times $d_w$.

To compute the packing in Fig. \ref{fig:compound}(c) we initially set $N_{max} = 12$ because this corresponds to an intuitive segmentation of the object, while we set $E = 50\%$. Then, we let the algorithm terminate at $N=N_{max}$ with an eventual packing efficiency of $E_{best} = 42\%$ that, for the specific object at hand, we reputed to be satisfactory.
Fig. \ref{fig:compound}(b) shows the results on the mosquito for which we set $N_{max} = 20$ and $E = 50\%$. These parameters led to a packing efficiency of $38\%$ and we decided to let the algorithm try up to a new $N_{max} = 30$. This produced a better packing $E_{best} = 48\%$ that we reputed to be satisfactory. For the robot model (Fig. \ref{fig:compound}(d)) we set $N_{max} = 10$, $E = 50\%$ and we obtained a satisfactory $E_{best} = 40\%$). For the rifle model (Fig. \ref{fig:compound}(a)) we set $N_{max} = 50$ and $E = 40\%$; with these parameters the algorithm terminated with $N=20$ parts that could be fit into a box according to the required target efficiency $E_{best} = E = 40\%$.

Detailed results are reported in Table \ref{tab:results}. We observe that the computing time is mostly dependent on the eventual number of parts that need to be packed. Also, we note that our size-driven ordering of the parts has a positive impact on the eventual packing efficiency; the only exception to this rule is the Rifle model, for which a random order has given a slightly better packing. Deactivating the ``placement into holes'' feature (Sec. \ref{sec:holes}) was not advantageous in any of our tests. Using 40 rotations instead of 10, or using 49 base sizes instead of 16, seems to produce a slight increase of the efficiency at the cost of a much longer computation time.

Comparing the whole split-and-pack algorithm to previous work is difficult because we are not aware of any other method that performs this operation (the recent concurrent work by Vanek and colleagues is \cite{vanek2014} discussed in the end of this section).
Its two fundamental algorithms that perform segmentation and packing, however, are worth a comparative assessment.

\begin{figure}
   \includegraphics[width=\linewidth]{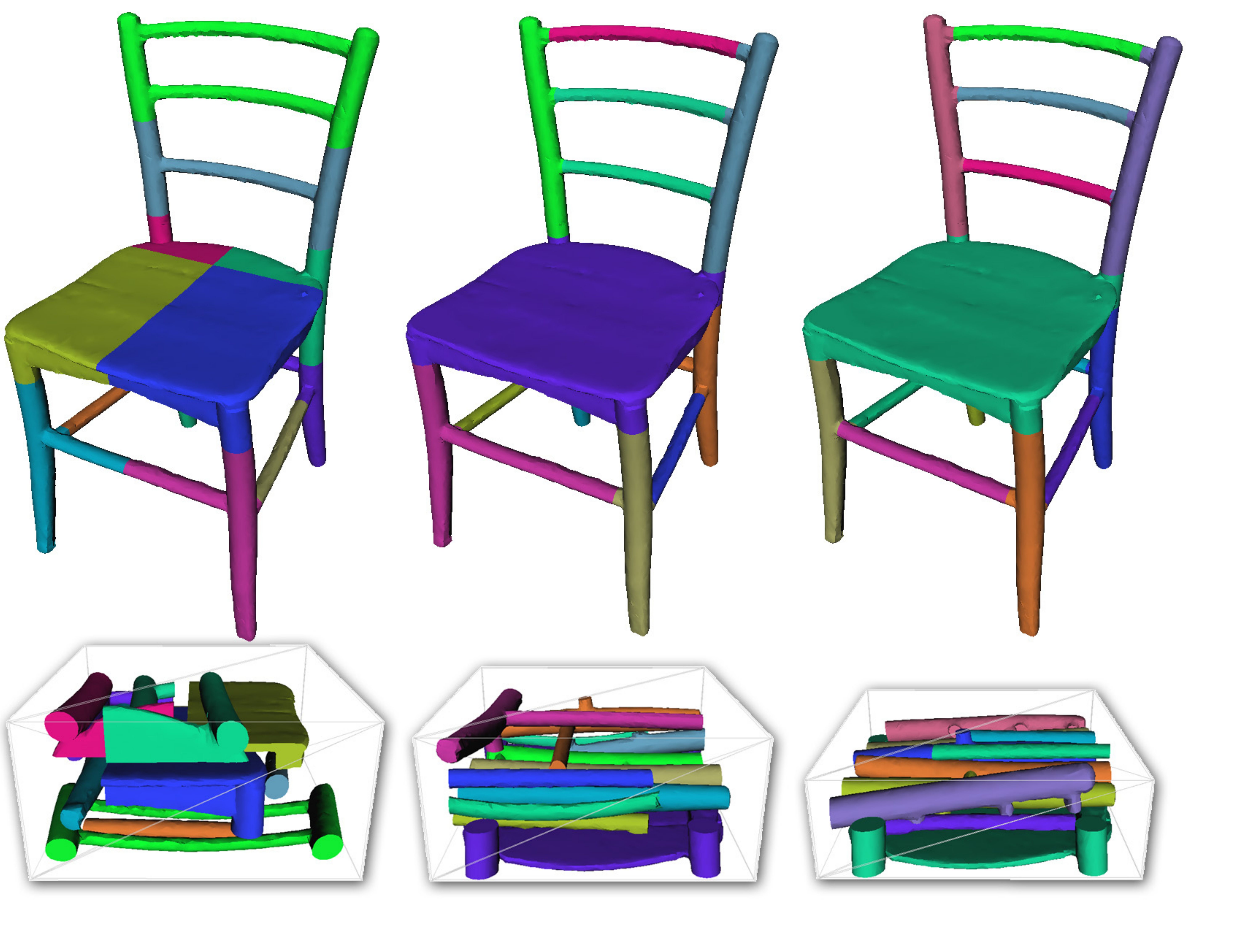}
   \caption{The 12 parts of the chair model subdivided as in [LBRM12] could be packed with an efficiency of 30\% (left). By setting $N_{max} = 12$ our algorithm could reach an efficiency of 39\%. If left to try up to $N_{max} = 14$, our algorithm could obtain an efficiency of 46\%.}
   \label{fig:choppercomp}
\end{figure}

Regarding the segmentation part, our conclusion is rather simple: the results of the new hierarchical segmentation based on the absolute aboxiness are very similar to those obtained in \cite{attene2008}.
This fact is not surprising because both the algorithms are driven by a volume difference: the only distinction is the type of \emph{container} used, which in this work is the part's minimum bounding box, while in \cite{attene2008} is the convex hull. This difference, however, is reflected in the \emph{semantic} quality of the resulting parts, which in our case is slightly worse (e.g. the mosquito's body is split in a non-intuitive way). Nontheless, we have repeated all our experiments using the method of \cite{attene2008} to produce the hierachies, and we observed that the eventual packing efficiency decreases by an average factor of 19\% (i.e. $(E_{Boxy} - E_{Convex}) / E_{Boxy} = 0.19$); if 40 rotations are used instead of 10, the efficiency gap becomes 17\% on average, which is still a significant difference.
Similarly, we have replicated some of the segmentations shown in \cite{luo2012} and computed a packing of the parts using our algorithm. By setting $N_{max}$ to the same number of parts used in \cite{luo2012} and $E = 100\%$, we could fairly compare the efficiency of such packings with the one obtained using our original aboxiness-based segmentation, and we observed that the latter exhibits an average increase of nearly 23\% (increase calculated as $(E_{ours}-E_{chopper})/E_{ours})$. In particular, we have tested the following models (see Fig. 18 in \cite{luo2012}):
$Fertility$ - 6 parts, $E_{chopper} = 22\%$, $E_{ours} = 33\%$;
$Octopus$ - 10 parts, $E_{chopper} = 28\%$, $E_{ours} = 33\%$;
$Hand$ - 7 parts, $E_{chopper} = 28\%$, $E_{ours} = 37\%$;
$Chair$ - 7 parts, $E_{chopper} = 30\%$, $E_{ours} = 39\%$.
For the $Chair$ model we also allowed the algorithm to perform two additional splits to achieve a significant $E_{ours} = 46\%$ (see Fig. \ref{fig:choppercomp}).

\begin{table*}[t!]
\centering
\caption{Packing results. $E_{best}$ reports the efficiency obtained with default parameters (i.e. 10 rotations and 16 base sizes in a $256^3$ voxel grid) while requiring to produce $N_{max}$ parts without limitations on the target efficiency. $E_{init}$ is the initial ratio between the object and its bounding box volume. $T$ is the time required by the whole split-and-pack process. All the other $Es$ represent the efficiency obtained by modifying either one default parameter or a functionality. Thus, $E_{rand}$ is what we obtain by replacing the size-driven placement order with a random order of the parts. $E_{noholes}$ = ``placement into holes'' feature is deactivated. $E_{R40}$ = 40 rotations are used instead of 10. $E_{B49}$ = 49 base sizes are used instead of 16 (growing factors from $+50\%$ to $-25\%$ with a step of $12.5\%$).}{
\begin{tabular}{ | l | l | c | c | c | c | c | c | c | c | c | c | }
\hline
\textbf{Model} 		&Fig.	   &$N_{max}$ &$E_{best}$ &$E_{init}$ &$T$ (s) &$E_{rand}$ &$E_{noholes}$ &$E_{R40}$ &$T_{R40}$ (s) &$E_{B49}$ &$T_{B49}$ (s)\\
\hline \hline
Naptune  	&\ref{fig:teaser}  	   &13   	&41 &5	  &211	    &37        &36           &42       &638              &42       &714 \\ \hline
Chair 1  	&\ref{fig:chairs}  	   &6   	&41	&9  	  &95	    &38        &37           &41       &280              &41       &288 \\ \hline
Dinosaur    &\ref{fig:dino}  	   &11   	&39 &6	  &187	    &39        &38           &41       &566              &41       &543 \\ \hline
Airplane   	&\ref{fig:cessna}  	   &8   	&38 &6	  &124	    &34        &38           &38       &368              &39       &375 \\ \hline
Chair 2   	&\ref{fig:choppercomp} &14   	&46 &4	  &207	    &38        &44           &46       &628              &46       &591 \\ \hline
Rifle   	&\ref{fig:compound}(a) &20   	&40 &12	  &306	    &41        &37           &40       &907              &42       &914 \\ \hline
Mosquito   	&\ref{fig:compound}(b) &30   	&48 &2	  &354	    &42        &44           &49       &1067             &49       &1104 \\ \hline
Monster   	&\ref{fig:compound}(c) &12   	&42 &5	  &191	    &36        &41           &43       &569              &42       &583 \\ \hline
Robot   	&\ref{fig:compound}(d) &10   	&40 &2	  &211	    &39        &40           &40       &633              &41       &641 \\ \hline
\end{tabular}}
\label{tab:results}
\end{table*}

\begin{figure}
   \centering
   \includegraphics[width=0.6\linewidth]{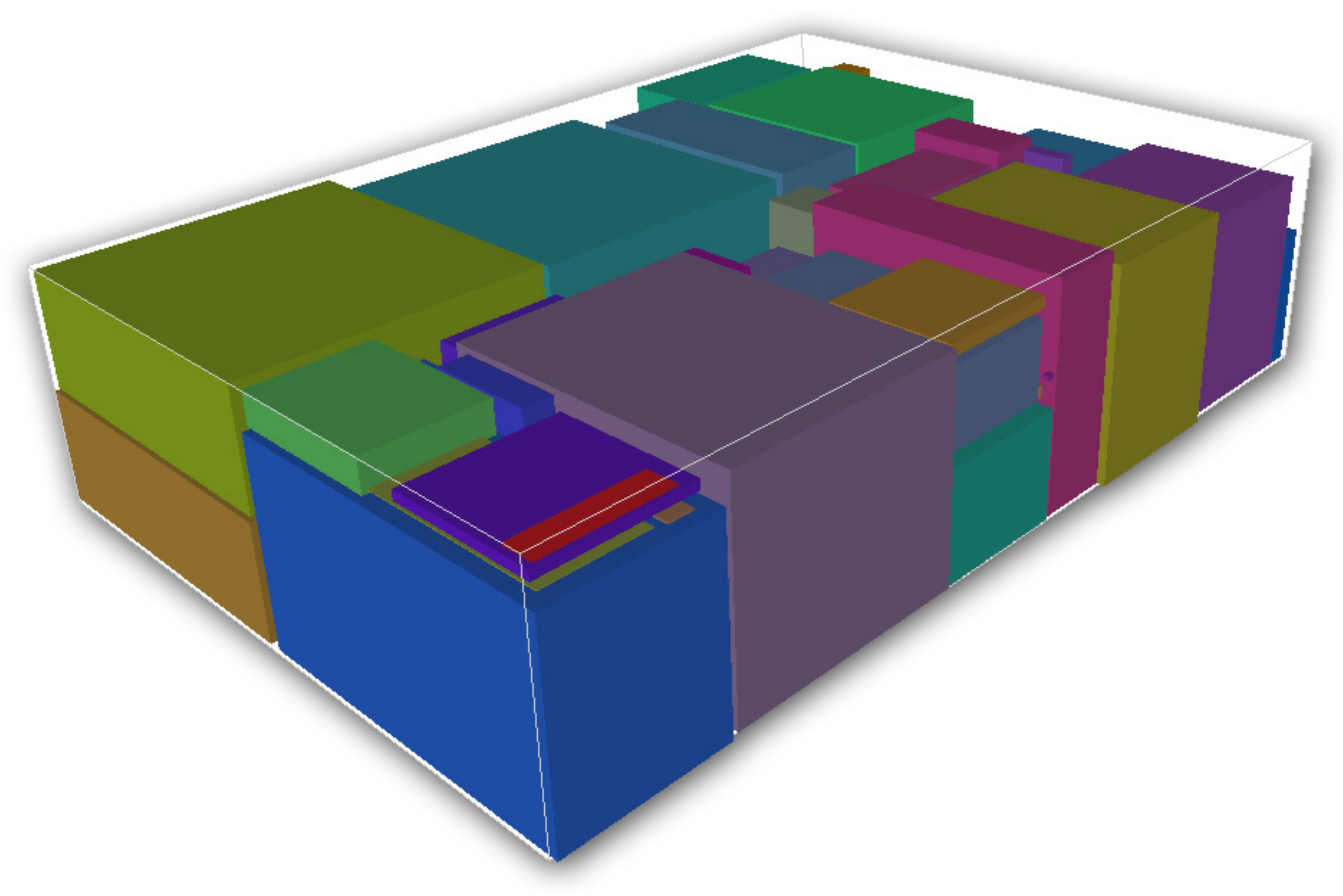}
   \caption{50 random boxes packed through our approach with an efficiency of 89\%.}
   \label{fig:boxes}
\end{figure}

\begin{figure}
   \centering
   \includegraphics[width=0.8\linewidth]{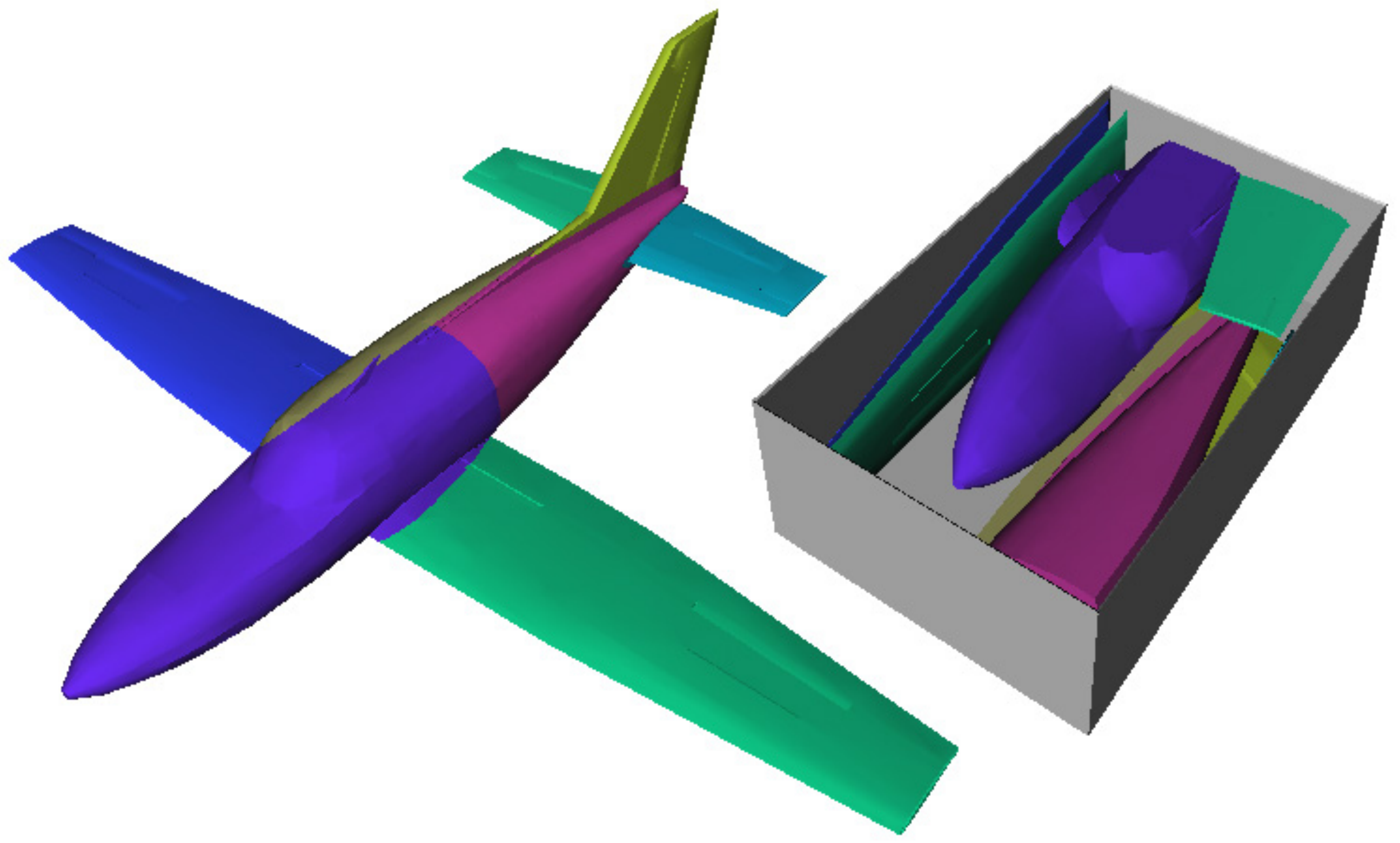}
   \caption{A model packed within its box without the top. This configuration is ready to be printed.}
   \label{fig:cessna}
\end{figure}

When compared with \cite{luo2012} our approach leads to parts which are more aligned with the object features with respect to BSP-driven cuts, and this has a positive impact on the aesthetics. Unfortunately, sometimes this comes at the cost of a less robust structure, but in most cases this is compensated by the use of the iron wire reassembly approach. Also, using \cite{luo2012} on meshes with long and thin features (e.g. the Mosquito in Fig. \ref{fig:compound}) is appropriate only if the printing volume is large enough to contain the largest thin feature (e.g. the Mosquito's legs). If it does not, any split would represent an issue for the creation of connectors. The same argument holds for models containing sheet-like features such as the aircraft wings in Fig. \ref{fig:cessna}.

As mentioned in Sec. \ref{sec:star_packing}, the only generic algorithm for single bin 3D packing \cite{chernov2010} seems to be unsuitable for arbitrarily-shaped objects as those treated in this paper. For the more specialized setting where only box-like objects (i.e. cartons) need to be packed, \cite{wu2010} appears to represent the state of the art. After having run 30 tests with randomly-generated boxes (see example in Fig. \ref{fig:boxes}), we have verified that the average packing efficiency achieved by our algorithm is roughly 88\%; thus, although our algorithm is more general, these results are comparable with the average 91\% reported in \cite{wu2010} for an analogous experiment.

Note that our choice of using $10$ rotations is motivated by the fact that this is exactly the number of mutually orthogonal orientations of a standard reference frame. Thus, this covers all the possibilities when parts are actual boxes, while it provides a good choice in all the other cases too because the parts are previously rotated so that their MBB becomes axis-aligned. 
Also, it is worth pointing out that the resulting packing efficiency varies significantly depending on the shape and relative size of the parts. On one hand, we may expect that equally-sized rectangular boxes can be fit with very high efficiency (theoretically, up to 100\%). Converesly, on the other hand we cannot expect too tight packings where the parts are arbitrarily shaped and exhibit a lot of variations in size. Regarding the size, however, our choice of using an absolute aboxiness for the segmentation partially tampers the problem because parts should not vary too much in size, but regarding the shape the problem is intrinsic, and a tight packing can be simply impossible unless the number of parts is allowed to grow significantly.

While this article was under review a quite similar work was published by another group \cite{vanek2014}. Such a concurrent algorithm, called $PackMerger$, aims at subdividing an object and packing the resulting parts, with the final goal of reducing the printing time and the support material required. Hence in \cite{vanek2014} the packing efficiency is not even reported whereas the material saving is measured in detail. In contrast, our goal is to produce small boxes to be shipped, and in this scenario maximizing the packing efficiency is fundamental whereas saving material is not a main concern. Thus, instead of trying to provide a quantitative comparison, we just outline how the two algorithms are expected to behave under various circumstances.

\begin{figure}
   \centering
   \includegraphics[width=.8\linewidth]{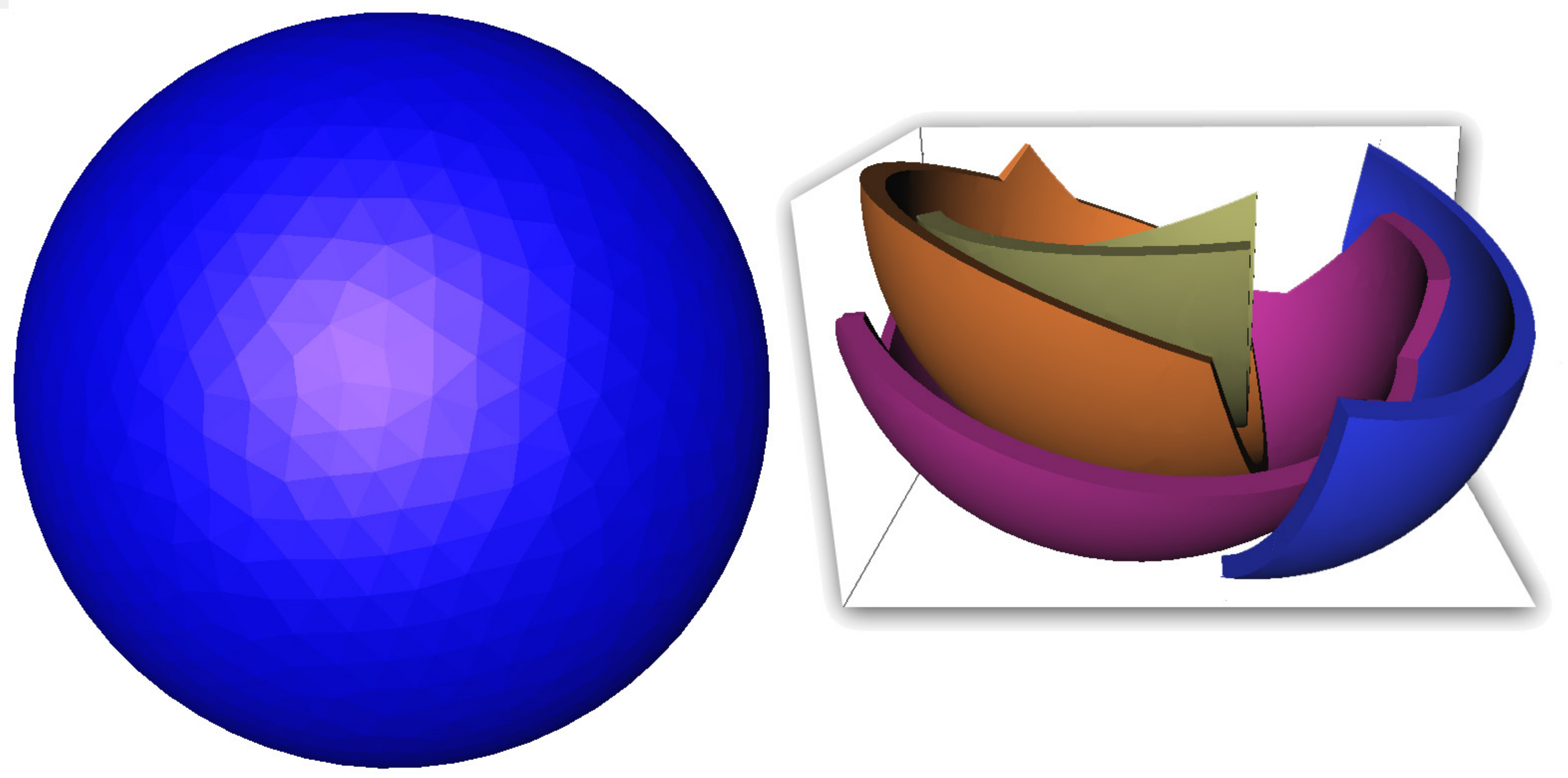}
   \caption{A hollow sphere packed by our algorithm to minimize the box volume.}
   \label{fig:hsphere}
\end{figure}

First, since the two methods have different objectives they produce different configurations. For example, in \cite{vanek2014} the pieces of a hollow sphere are stacked one on top of the other because the algorithm strives to minimize the use of support material, whereas since our method focuses on the packing efficiency it may produce different arrangements (Fig. \ref{fig:hsphere}).
Related to this, it is worth considering that $PackMerger$ first turns the input solid to a shell by hollowing its inner parts. This feature has both advantages and drawbacks: the printed object requires less material, the parts are generally thinner and can be packed within smaller volumes. On the other hand, since the gluing areas are narrower, reassembling the object is trickier and the result is structurally weaker.
Also, in $PackMerger$ one of the terms of the cost to be minimized controls the amount of support material. By increasing that term's weight in the cost equation one can actually reduce the use of support materials, but this comes at the cost of a lower packing efficiency.

In our split-and-pack method, when the efficiency is not sufficient one specific part is split. In $PackMerger$ an opposite approach is employed, the algorithm starts with an over-segmented object and then tries to merge adjacent parts. The two parts to be merged are not pre-determined and can change while iterating, which leads to a higher flexibility and potentially tighter packings. Even in this case, however, choosing among more possibilities requires the packing algorithm to iterate more times, and this leads to longer computations. 
Regarding the assemblability of the resulting parts, $PackMerger$ prevents the creation of too small split areas through a pre-merging process. This is not done in our approach, though the problem is mitigated by the use of iron wires. On the other hand, our method provides a refinement of the split areas and a reassembly order that guarantees a feasible reconstruction.

\subsection{Limitations}
Besides the obvious limitations that a sub-optimal algorithm necessarily has, our work is still limited in its actual applicability for some 3D printing applications.
While printing the box using fused deposition, standard print paths typically produce long plastic filaments that tend to shrink as they cool, and in some cases this effect may lead to print failures. Thus, standard plastic such as ABS may become inappropriate to print our packed shapes, whereas less deformable plastic (e.g. PLA) is necessary to avoid failures. Also, in some cases the split areas separating two parts can be extremely small, and this makes the eventual reassembly a delicate operation even through the iron wire approach. Iron wire makes the reassembly reasonably robust for most applications, but our algorithm may be not appropriate when the resulting objects have strong robustness requirements (e.g. a chair that must sustain a person weight): when such a robustness is an issue, an integration with existing approaches such as \cite{stava2012} becomes necessary. Finally, the use of a greedy approach to segment the object combined with the cost in equation \ref{eq:aboxiness} may occasionally lead to unintuitive segmentations (e.g. the fuselage in Fig. \ref{fig:cessna}) that have a negative impact on the aestetics of the reassembled model.

Finally, since some packing configurations can be mutually interlocking \cite{song2012}, and since small parts can be placed within larger hollow parts without the possibility to be extracted, it is important to avoid these particular arrangements. In principle, one can simply deactivate the ``placement into holes'' functionality to guarantee that all the parts can be extracted from the pack, but in general this would have a negative impact on the packing efficiency. In any case, we argue that our packings are hardly interlocking because they do not fill the space tightly enough, but the user should bear in mind this risk when hollow objects are being split and packed.

\section{Conclusions and future work}
This paper demonstrates that an arbitrary 3D shape can be effectively split and packed based on an algorithmic approach, even if the computation of an optimal solution is NP-complete. Though relevant previous work do not provide satisfactory solutions to this problem, some approaches exist for the treatment of specific shapes such as boxes or synthetic furniture pieces. We concluded that our packing approach produces results which are comparable with these previous methods while being much more general in its applicability.

Being a first attempt to solve a so-generic packing problem, our algorithm is still far from optimality and there is plenty of room for improvements and further research. We have split the problem in two successive phases of hierarchical segmentation and packing, but we argue that better results are possible if these two steps are merged into a more integrated algorithm where the definition of the part shapes, and not just their number, is driven by their actual possibility to be packed, possibly by considering the current configuration of the previously-placed parts.
Furthermore, we suspect that further optimizations can be achieved by exploiting rotation-invariant local shape descriptors and geometric complementarity \cite{daras2011} to identify potentially good rotations without the need to blindly iterate among numerous pre-defined orientations.

\section*{Acknowledgements}
This work is partly supported by the EU FP7 Project N. ICT-2011-318787 (IQmulus) and by the international joint project on \emph{Mesh Repairing for 3D Printing Applications} funded by Software Architects Inc (WA, USA). Thanks are due to the SMG members at IMATI for helpful discussions.

\bibliographystyle{eg-alpha-doi}
\bibliography{packed_shapes}

\end{document}